\documentstyle[aps,prl,multicol,epsf,epsfig,float]{revtex}

\begin{document}
\newcommand{\ri}{{\rm i}}
\newcommand{\re}{{\rm e}}
\newcommand{\bx}{{\bf x}}
\newcommand{\br}{{\bf r}}
\newcommand{\bk}{{\bf k}}
\newcommand{\bn}{{\bf n}}
\newcommand{\rg}{{\rm g}}
\newcommand{\tr}{{\rm tr}}
\newcommand{\xmax}{x_{\rm max}}
\newcommand{\ra}{{\rm a}}
\newcommand{\rx}{{\rm x}}
\newcommand{\rs}{{\rm s}}
\newcommand{\rP}{{\rm P}}
\newcommand{\up}{\uparrow}
\newcommand{\down}{\downarrow}
\newcommand{\hc}{H_{\rm cond}}

\sloppy

\title{Creation of Entanglement by Interaction with a Common Heat Bath}
\author{Daniel Braun}
\address{c/o
IBM T.J. Watson Research Center, Yorktown Heights, NY 10598}
\maketitle
\centerline{\today}
\begin{abstract}
\begin{center}
\parbox{14cm}{I show that entanglement between two
qubits can be generated if the two qubits interact with a common heat bath in
thermal equilibrium, but do not interact directly with each other. 
In most situations the entanglement is created for a very short time after
the interaction with the heat bath is switched on, but depending on system,
coupling, and heat bath, the entanglement may
persist for arbitrarily long times. This mechanism sheds new light on the
creation of entanglement. A particular example of two quantum
dots in a closed cavity is discussed, where the heat bath is given by the
blackbody radiation.
}
\end{center}
\end{abstract}

\begin{multicols}{2}

Since the discovery of quantum mechanics, ``entanglement'' has been
considered a hallmark of quantum behavior
\cite{Schroedinger35}. Two quantum systems $A$ and $B$  in a pure state are
called entangled, if their quantum mechanical state vector  $|\psi\rangle$
can {\em not} be written as product of two states $|\phi_A\rangle$ and
$|\phi_B\rangle$ in the Hilbert spaces of $A$ and $B$, respectively. 
The last few years of research on quantum information processing have lead
to the picture of entanglement as a precious resource. Entanglement
plays an important role in 
super--dense coding \cite{Bennett92} and
quantum  teleportation \cite{Bennett93}, and 
is necessary for the exponential
speed--up  of quantum algorithms compared to classical
algorithms \cite{Jozsa02}.   

Recently
investigated examples of the controlled creation of entanglement include
trapped  
ions that interact electrostatically (or more precisely exchange
phonons in a chain of ions
\cite{Cirac95}), and the 
entanglement of atoms in a cavity by the
interaction with a specific electromagnetic mode of the cavity
\cite{Haroche95,Beige00}. In the latter example entanglement can be created
even in the case where the cavity mode is itself coupled to many more
degrees of freedom of the electromagnetic environment, i.e.~if the cavity is
more or less leaky. Nevertheless, in all these examples a
third system with one or few degrees of 
freedom is clearly singled 
out by mediating the interaction between the atoms or ions. This is true even
for strongly leaking cavities capable of supporting 
super--radiance \cite{Bonifacio71}, which may still entangle
atoms \cite{Schneider01}. 

In the following I show that entanglement can be created if the two systems
interact neither directly, 
nor with a third system with only one or a few singled out degrees of freedom,
but interact with the (possibly infinitely many) degrees of freedom of a
heat bath in 
thermal equilibrium. This is a priori not obvious since interactions with a
heat bath lead typically to 
very rapid decoherence \cite{Zurek81}, thus to classical states and
the destruction of quantum entanglement. 
Indeed, we will see that the entanglement created may die again on a
decoherence time scale of the 
system. However, notable exceptions exist: {\em i}) if the two systems
are coupled in a symmetric way to the environment the entanglement will be
protected \cite{Zurek82} --- much in the spirit of what is known from
coherent rotational tunneling 
\cite{Stevens83}, long living Schr\"odinger cat states
\cite{Braun98.2} 
and decoherence free subspaces  \cite{Beige00,Lidar98,Braunbook01}.  
{\em ii}) Many environments will lead, for systems with degenerate energy
levels and in sufficiently high dimension, to incomplete 
decoherence, a surprising effect to be discussed below.

When dealing with a ``heat bath'', i.e.~another system with very many
degrees of freedom over which we do not have microscopic control,
the definition of entanglement has to be generalized to mixed states.
A state of a bipartite system is said to be ``separable'', iff the density
matrix of the state can be written as
\begin{equation} \label{rhosep}
\rho=\sum_{i=1}^N p_i \rho^A_i\otimes \rho^B_i\,,
\end{equation}
where the $p_i$ are probabilities ($0\le p_i\le1$), $\rho^A_i$ and
$\rho^B_i$ are density matrices for the two subsystems $A$ and $B$, and $N$
is an arbitrary integer. A state
that is not separable is said to be entangled \cite{Nielsen00}.
 A simple criterion for entanglement of bipartite systems of
dimensions 
$2\times 2$ or $2\times 3$ was proven by the Horodecki family
\cite{Horodecki96}: A state $\rho$ of a
$2\times 2$ or $2\times 3$ bipartite system is separable, iff $\rho$ has a
non--negative partial transpose $\rho^{T_A}$. The partial transpose
$\rho^{T_A}$ is obtained by transposing in a matrix
representation of $\rho$ only the indices corresponding to subsystem $A$,
i.e.~$\rho^{T_A}_{ik,jl}=\rho_{jk,il}$ with $ \rho_{jk,il}=\langle j|\otimes
\langle k|\rho|i\rangle \otimes |l\rangle$. 

Suppose Alice and Bob both own a qubit with basis states $|0\rangle$ and
$|1\rangle$ over which they 
have local control.  The qubits do not interact with
each other. Thus, the Hamiltonian representing the two qubits is simply
$H_{AB}=H_A+H_B$, where $H_A$ acts only on Alice' Hilbert space, and $H_B$
only on 
Bob's. Suppose further that the qubit states $|0\rangle$ and $|1\rangle$ are
energy eigenstates with degenerate energies, for both qubits.  In this case,
$H_{AB}=0$ up to an irrelevant constant. For situations where the degeneracy
is not exact, let us assume at least that 
the inverse level spacing is much larger than any time scale that we are
interested in. The dynamics induced by a finite
$H_A$ or $H_B$ can then  be neglected and we can drop the ``system
Hamiltonian'' $H_{AB}$ \cite{Braun01}. 
The heat bath will be
described as a 
collection of $N$ harmonic oscillators,
\begin{equation} \label{Hbath}
H_{\rm bath}=\sum_{i=1}^N\left(\frac{1}{2m}p^2_i+\frac{1}{2}m\omega_i^2q_i^2\right)\,.\end{equation}
For the interaction with the heat bath we assume a coupling Hamiltonian
\begin{equation} \label{Hint}
H_{\rm int}=(S^A+S^B)B\,,\,\,\,\,\,\,B=\sum_{k}g_k q_k\,,
\end{equation}
where $S^A$ and $S^B$ are ``coupling agents'' acting on the Hilbert spaces
of Alice and Bob, respectively, and the $g_k$ are coupling constants to the
$k$th oscillator. An example that is described by
(\ref{Hint}) will be analyzed in detail below. 

Let us further assume
that the qubit basis states $|0\rangle$ and $|1\rangle$ are eigenstates of
$S^A$ and $S^B$ with eigenvalues $a_0$, $a_1$, and $b_0$,
$b_1$, respectively. The combined computational
basis states $|00\rangle$, $|01\rangle$, $|10\rangle$, and $|11\rangle$ are
then eigenvectors of $S^A+S^B$ with corresponding eigenvalues $a_0+b_0$,
$a_0+b_1$, $a_1+b_0$, and  $a_1+b_1$,
respectively. 

Protecting their qubits momentarily from the
environment, Alice and Bob 
prepare pure initial states $|\varphi^A\rangle$ and $|\varphi^B\rangle$ of
their respective qubits. The heat bath is assumed to 
be initially in thermal equilibrium at temperature $T$, and so the total
initial state is the density matrix 
\begin{equation} \label{W0}
W(0)=|\varphi^A\rangle\langle\varphi^A|\otimes
|\varphi^B\rangle\langle\varphi^B|\otimes \frac{1}{Z}\re^{-H_{\rm bath}/k_{\rm B}T}\,,
\end{equation}
where $Z=\tr_{\rm bath}\re^{-H_{\rm bath}/k_{\rm B}T}$ and $k_{\rm B}$
denotes Boltzmann's 
constant. 
The time evolution of Alice' and Bob's qubits alone is described by 
the reduced density matrix $\rho(t)=\tr_{\rm bath}W(t)$. That time
dependence was 
calculated for an arbitrary system with negligible system
Hamiltonian and coupled as in (\ref{Hint})
to a heat bath of harmonic oscillators in
\cite{Braun01}. The result can be phrased in terms of two 
functions $f(t)$ and $\varphi(t)$, 
\begin{eqnarray}
f(t)&=&\sum_k\frac{g_k^2(1+2\overline{n}_k)}{2m\hbar
\omega_k^3}\left(1-\cos\omega_k t\right)\,,\nonumber\\
&=&{\rm Re}\frac{1}{\hbar^2}\int_0^t
ds\,s C(t-s)\\
\varphi(t)&=&\sum_k\frac{g_k^2}{2m\hbar
\omega_k^2}\left(t-\frac{\sin\omega_k t}{\omega_k}\right)\,,\nonumber\\
&=& {\rm Im}\frac{1}{\hbar^2}\int_0^t ds\,s C(t-s)
\end{eqnarray}
where $\overline{n}_k$ denotes the thermal occupation of the $k$th mode and $C(t)=\langle B(t)B(0)\rangle$ represents the thermal bath correlation
function. In
the basis of eigenstates of $S^A+S^B$ (the ``pointer 
basis'' \cite{Zurek81}), the time evolution of 
$\rho_{ij,kl}(t)$ ($i,j,k,l=0,1$) is given by
\begin{eqnarray} 
\rho_{ij,kl}(t)&=&\rho_{ij,kl}(0)\exp\Big(-\big(a_i+b_j-a_k-b_l\big)^2f(t)\\\label{rhot} 
&&+\ri 
\big((a_i+b_j)^2-(a_k+b_l)^2\big)\varphi(t)\Big)\,.\nonumber
\end{eqnarray}
In general this time evolution leads to a rapid decay
of the off--diagonal matrix elements --- unless $S^A+S^B$ has degenerate
eigenvalues.

Suppose Alice and Bob prepare
the  initial states $|\phi^A\rangle=(|0\rangle-|1\rangle)/\sqrt{2}$ and 
$|\phi^B\rangle=(|0\rangle+|1\rangle)/\sqrt{2}$, i.e.
\begin{equation} \label{rho0}
\rho(0)=\frac{1}{4}\left(
\begin{array}{rrrr}
1&1&-1&-1\\
1&1&-1&-1\\
-1&-1&1&1\\
-1&-1&1&1
\end{array}
\right)\,.
\end{equation}
and assume for the moment the symmetric coupling situation 
$a_0=b_0=0$, and 
$a_1=b_1=1$, absorbing eventual prefactors into the coupling constants
$g_k$. It is straight forward to compute numerically the eigenvalues of
the 
partially transposed density matrix $\rho^{T_A}$ for given $f(t)$ and
$\varphi(t)$. Note that these functions vanish at $t=0$ and are strictly
positive for times $t>0$; for small $t$ ($\omega_k t\ll 1$)
always {\em both} $f(t)$ and $g(t)$ become finite, with $f(t)\propto
\varphi(t)^{2/3}$. By parameterizing the eigenvalues directly by $f(t)$
and $\varphi(t)$ one can examine all possible (harmonic) heat baths at the
same time. A given heat bath leads to a certain path in the $f,\varphi$
plane. Fig.\ref{fig.ew0}  shows the smallest eigenvalue $\lambda_0$ of
$\rho^{T_A}$ as 
function of $f(t),\varphi(t)$. The eigenvalue is zero for $t=0$, where
both $f$ and $\varphi$ vanish: since the two qubits were
prepared in a product state, the partial transpose is the same as the
original matrix, and the Schmidt decomposition gives one eigenvalue
unity and three equal zero. As soon as $f(t)$
and $\varphi(t)$ aquire a finite value,  $\lambda_0$ becomes negative,
however, meaning that the two qubits get entangled.  For larger values of $f$
and $\varphi$, the absolute value of $\lambda_0$ decays again, and 
asymptotically, for $f(t)\to\infty$, the state 
\begin{equation} \label{rhoinf}
\rho_{\infty}=\frac{1}{4}\left(
\begin{array}{rrrr}
1&0&0&0\\
0&1&-1&0\\
0&-1&1&0\\
0&0&0&1
\end{array}
\right)\,
\end{equation} 
is reached, independent of the behavior of the imaginary
part. Alice's  partial transpose $\rho_\infty^{T_A}$ of this matrix has 
eigenvalues 1/2, 1/4 (doubly degenerate) and 
zero, so that for $f(t)\to\infty$ no entanglement is left. Using second order
perturbation theory in the deviation of $\rho(t)^{T_A}$ from
$\rho_\infty^{T_A}$ one easily shows,
however, that for all arbitrarily large but finite $f(t)$ the two qubits stay
entangled. 
  
Finite entanglement is created at short times also in the case that
$a_0+b_1$ and $a_1+b_0$ are {\em not} degenerate. Numerical investigation
shows that for a given $f(t)$ and
$\varphi(t)$ the positivity of $\rho^{T_A}$ may be even more
strongly violated for non--perfect degeneracy.
Non--perfect degeneracy
changes things drastically, however, for large $f(t)$ when for all finite
deviations from degeneracy the non--entangled state
$\rho_{\infty}=\frac{1}{4}{\rm 
diag}(1,1,1,1)$ is  reached. Therefore there will be a {\em finite}  time
after which the 
initial state becomes separable --- if $f(t)$ reaches large values. 

It is easy to show numerically that the choice of the initial state is not
crucial. As long as both states contain components of both $|0\rangle$
and $|1\rangle$, the heat bath creates entanglement between the two qubits.
And I have also checked that the  interaction with a common heat bath
can create entanglement between a qubit and a qutrit (i.e.~a 
$2\times 3$ bipartite system).
\noindent
\begin{minipage}{9cm}
\begin{center}
\begin{figure}[h]
\epsfig{file=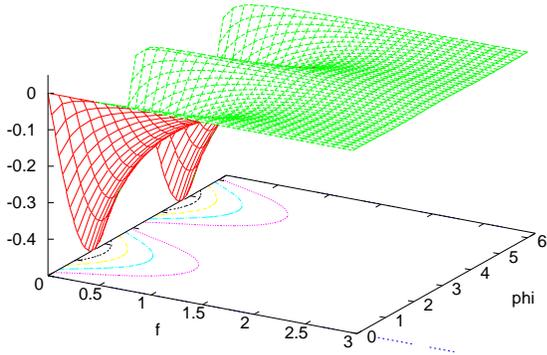,width=9cm} \\[0.2cm]
\caption{Smallest eigenvalue of $\rho^{T_A}$ as a function of $f$ and $\varphi$}\label{fig.ew0}  
\end{figure}
\end{center}
\end{minipage}
\vspace*{0.1cm}

Let me finally propose a concrete system where the effect might in principle
be observable. Consider two double--well quantum dots enclosed in an ideally
conducting, box--shaped cavity, with edge dimensions $a$, $b$, and $c$ in
$x$, $y$, and $z$ directions, respectively. The two
quantum dots are assumed identical, with the two--dimensional
electron gas in the $y=b/2$ plane, and with two identical wells to the
right and left (in $z$ direction) each, which might be electrostatically
defined by suitable gate--electrodes. 
The symmetry centers of the double--well quantum dots are located at
$\bx_A=(a/4,b/2,c/2)$ and $\bx_B=(3a/4,b/2,c/2)$ for dot $A$ and $B$,
respectively. The centers of the wells are separated by a distance $d$ and
we will assume that there exist in both dots two states $|0\rangle$ and
$|1\rangle$, localized in the right and left well, such that they are
eigenstates of the dipole operators, i.e. 
$e \br_A$, $\langle 0|\br_A|0\rangle=-d/2 (0,0,1)=-\langle
1|\br_A|1\rangle$, $\langle 0|\br_A|1\rangle=0$ for dot $A$, where $e$ is
the electron charge and $\br_A=\bx-\bx_A$ is the position of an electron
with respect to the center of the dot. This can be   achieved to
very good approximation by a very high barrier between the two wells, which
leads to exponentially small overlap of the two wave functions, and
negligible tunneling splitting. For dot $B$
the two states are chosen in the opposite way, $\langle
0|\br_B|0\rangle = d/2 (0,0,1)=-\langle 1|\br_B|1\rangle$. The cavity
supports TE and TM modes 
(``transversal'' relative to the arbitrarily chosen $z$--direction as
propagation direction). For
the above geometry the two dots interact only with the TM modes, if we
describe 
the interaction between the dots and the electromagnetic field in dipole
approximation. This is suitable for
temperatures where only modes with wave lengths much larger than $d$ are
populated, i.e.~$k_B T\ll 2\pi \hbar c_0/d$ ($c_0$ is the speed of light).  
The interaction in dipole approximation reads
\begin{eqnarray} 
H_{\rm int}^{A,B}&=&-\sum_{i,j=0}^1 P_{i,j}^{A,B}|i\rangle\langle j| \sum_{\bk} {\cal
E}_{\bk}(\bx_{A,B}) \hat{q}_{\bk}\,,\label{Hdip}
\end{eqnarray}
where $P_{i,j}^{A,B}=e \langle i|\br_{A,B}| j\rangle$ are the dipole
matrix elements defined above, ${\cal
E}_{\bk}=\sqrt{\frac{m}{2
\mu_0}}(\varphi_x,\varphi_y,\varphi_z)$
(with $\epsilon_0$ and $\mu_0$ electric permeability and magnetic
susceptibility of vacuum, in SI units), and $\hat{q}_\bk$ is the electric field
amplitude of mode $\bk$ with the dimension of a length, chosen as coordinate
of the harmonic oscillator in the 
quantization of the field \cite{Scully97}. The mass $m$
introduced formally for this purpose will cancel out again in 
the final expressions for $f(t)$ and $\varphi(t)$.   
The functions
$\varphi_x$, $\varphi_y$, and $\varphi_z$ define the spatial structure of
the modes. Here, only $\varphi_z$ is needed, which for perfectly conducting
walls is given by 
\cite{Jackson75} 
\begin{equation} \label{phiz}
\varphi_z(\bx)=-\frac{2\sqrt{2}}{\sqrt{V}}\frac{k_\perp}{\epsilon_0}\sin(k_x x)\sin(k_y y) \cos(k_z z)\,;
\end{equation}
$V=abc$ denotes the total volume, $k_\perp=\sqrt{k_x^2+k_y^2}$ are
the transverse wave numbers, and the wave vector is given by
$(k_x,k_y,k_z)=\pi (m/a,n/b,p/c)$, $m,n,p=0,1,2,\ldots$. Thus,
\begin{eqnarray} 
H_{\rm int}^{A,B}&=&\mp \sigma_x^{A,B}\sum_{\bk} g_{\bk}^{A,B}
\hat{q}_{\bk}\,,\\\label{HintAB}
 g_{\bk}^{A}&=&e d \sqrt{\frac{m}{\mu_0 V}}\frac{k_\perp}{\epsilon_0}\sin(k_x\frac{a}{4})\sin(k_y\frac{ b}{2})\cos(k_z\frac{c}{2})\label{ga}\\
 g_{\bk}^{B}&=&e d \sqrt{\frac{m}{\mu_0 V}}\frac{k_\perp}{\epsilon_0}\sin(k_x\frac{3a}{4})\sin(k_y\frac{ b}{2})\cos(k_z\frac{c}{2})\,.\label{gb}
\end{eqnarray} 
The upper sign refers to dot $A$, the lower to $B$, and the operators
$\sigma_x^{A,B}$ are defined for both systems as $\sigma_x=|0\rangle\langle
0|- |1\rangle \langle 1|$. One easily sees that the modes with even $m$
couple to 
$\sigma_x^A+\sigma_x^B$, those with odd $m$ couple to
$\sigma_x^A-\sigma_x^B$. However, the odd modes can be suppressed by a very
thin, uncharged, and perfectly conducting wire in the $z$ direction along
the $x=a/2,y=b/2$ axis 
of the cavity, since they  have non--vanishing tangential electrical field at the position
of the wire. We then obtain the coupling
Hamiltonian (\ref{Hint}) with $S^A=\sigma_x^A$, $S^B=\sigma_x^B$, 
$g_\bk=-g_\bk^A$, and quantized wave vectors
$\bk=\pi((4n_x+2)/a,(2n_y+1)/b,2n_z/c)$, $n_x,n_y,n_z=0,1,2,\ldots$. 
The resulting expressions for $f(t)$ and $\varphi(t)$ are divergent, and the
sum over $\bk$ needs a cut--off. For a cavity made out of a real
metal a natural cut--off frequency is the plasma frequency $\omega_p$,
since the metal looses its reflectivity for $\omega>\omega_p$
\cite{Kittel86}. 
Converting the sums over $\bk$ into integrals, defining
$\zeta(d,\beta)=e^2d^2\mu_0/(\pi^2c_0\hbar^3\beta^2)$, and $\tau=\beta\hbar/2$,
one finds  
$f(t)=\zeta(d,\beta)\tilde{f}(t)$ and 
$\varphi(t)=\zeta(d,\beta) \tilde{\varphi}(t)$ with
\begin{eqnarray}
\tilde{f}(t)&=&\frac{1}{4}\int_0^\infty dx\,x\coth(x)\,c(x)
\left(1-\cos\frac{xt}{\tau}\right)\label{tildef}\\ 
\tilde{\varphi}(t)&=&\frac{1}{3}\int_0^\infty dx\,x^2c(x)\left(\frac{xt}{\tau}-\sin\frac{xt}{\tau}\right)\,,\label{tildev} 
\end{eqnarray}
where $c(x)$ is a cut-off function (to be specific, say $c(x)=\exp(-x/x_{\rm
max})$) 
For an aluminum cavity, $\hbar\omega_p=15.3$eV \cite{Kittel86}, and we have,
at $T=100$mK, $\tau\simeq 3.8\,10^{-11}$s and $x_{\rm
max}=\omega_p\tau\simeq 8.8\, 
10^5$. The main contribution to the integrals therefore stems from $x\gg 1$,
where we can approximate the coth--function by 1. For the
exponential cut--off we then have
\begin{equation} \label{tildef2}
\tilde{f}(t)=\xmax^2\left(1-\frac{\cos\left(2 \arctan (t
\xmax/\tau)\right)}{1+t^2\xmax^2/\tau^2}\right)\,, 
\end{equation}
a function that saturates for $t\xmax \gg \tau$ at the value 
$\xmax^2$, after reaching a maximum at $t/\tau\simeq \xmax^{-1}$ . For
quantum dots with 
$d=10$nm, $T=100$mK, $\zeta(d,\beta)\simeq 1.8\,10^{-15}$, and $f(t)$
reaches a maximum after a time of the order 
$10^{-17}$s before saturating at $f(t)\simeq 0.0014$. This means that
decoherence remains incomplete even for non--symmetric couplings, 
and the entanglement created by the interaction with the heat bath is
preserved, till other 
decoherence mechanisms neglected in the above analysis kick in. Note that
such incomplete decoherence is a rather general result for systems with
degenerate energy levels. In fact, by
integrating the time dependent part in eq.(\ref{tildef}) from zero to
$t$ one obtains for $t\to\infty$ a Dirac delta function at $x=0$, and the
remaining 
integral over $x$ will give a finite constant. Thus, the time dependent part
of $\tilde{f}(t)$ has to 
decay for $t\to\infty$ faster than $1/t$, leaving the time independent part
$\int  
dx\,x\coth(x)c(x)/4$. Note that the factor $x$, the spectral weight of the
heat bath at zero 
frequency, is essential in this
reasoning. Decoherence will always remain incomplete (in the sense of finite 
$f(t)$ for $t \to \infty$, depending on the circumstances even $f(t)\ll 1$)
between degenerate energy levels for spectral 
weights that vanish at zero frequency faster than the first power of the
frequency. 

The presented scheme has an advantage over
conventional creation of entanglement if Alice and Bob are so far
apart that a direct interaction is difficult to achieve. Since $f(t)$ and
$\varphi(t)$ do not depend on the volume of the cavity,
very large cavities should be possible
with corresponding large distances between Alice and Bob; using the
non--symmetric coupling scheme one might even envisage to dispose of the
cavity altogether and rely on the long wave--length continuum of the cosmic
electromagnetic background radiation to create entanglement between very
remote quantum 
dots.  While only a small amount of mixed state entanglement will be
created, it is well known that all entanglement of a $2\times
2$ bipartite system can be distilled into a pure entangled state
\cite{Horodecki97}, given sufficiently many realizations of the input states
and local coherent control.

As a conclusion I have shown that entanglement can be created between two
qubits that interact solely with a common heat bath with very many degrees of
freedom. The explicit example of two double--well quantum dots in a
cavity was calculated, and the phenomena of ``incomplete decoherence'' was
revealed, which may, as much as symmetric couplings to the environment,
preserve the entanglement created by the heat bath. 

{\em Acknowledgements:} I would like to thank Fritz Haake and Walter Strunz
for many discussions on decoherence.

\end{multicols}
\end{document}